\documentclass[twocolumn,aps, prb, showpacs,preprintnumbers,amsmath,amssymb,a4paper,superscriptaddress]{revtex4} 

\usepackage{graphicx,subfigure}
\usepackage{dcolumn}
\usepackage{bm}
\usepackage[percent]{overpic}
\usepackage{wasysym}
\usepackage{longtable}

\begin{document}
\newcommand{\Arg}[1]{\mbox{Arg}\left[#1\right]}
\newcommand{\bb}{\mathbf}
\newcommand{\braopket}[3]{\left \langle #1\right| \hat #2 \left|#3 \right \rangle}
\newcommand{\braket}[2]{\langle #1|#2\rangle}
\newcommand{\be}{\[}
\newcommand{\br}{\vspace{4mm}}
\newcommand{\bra}[1]{\langle #1|}
\newcommand{\braketbraket}[4]{\langle #1|#2\rangle\langle #3|#4\rangle}
\newcommand{\braop}[2]{\langle #1| \hat #2}
\newcommand{\dd}[1]{ \! \! \!  \mbox{d}#1\ }
\newcommand{\DD}[2]{\frac{\! \! \! \mbox d}{\mbox d #1}#2}
\renewcommand{\det}[1]{\mbox{det}\left(#1\right)}
\newcommand{\ee}{\]} 
\newcommand{\eg}{\textbf{\\  Example: \ \ \ }}
\newcommand{\Imag}[1]{\mbox{Im}\left(#1\right)}
\newcommand{\ket}[1]{|#1\rangle}
\newcommand{\ketbra}[2]{|#1\rangle \langle #2|}
\newcommand{\kp}{\arccos(\frac{\omega - \epsilon}{2t})}
\newcommand{\ldos}{\mbox{L.D.O.S.}}
\renewcommand{\log}[1]{\mbox{log}\left(#1\right)}
\newcommand{\Log}{\mbox{log}}
\newcommand{\Modsq}[1]{\left| #1\right|^2}
\newcommand{\nb}{\textbf{Note: \ \ \ }}
\newcommand{\op}[1]{\hat {#1}}
\newcommand{\opket}[2]{\hat #1 | #2 \rangle}
\newcommand{\occ}{\mbox{Occ. Num.}}
\newcommand{\Real}[1]{\mbox{Re}\left(#1\right)}
\newcommand{\so}{\Rightarrow}
\newcommand{\sol}{\textbf{Solution: \ \ \ }}
\newcommand{\thetafn}[1]{\  \! \theta \left(#1\right)}
\newcommand{\tin}{\int_{-\infty}^{+\infty}\! \! \!\!\!\!\!}
\newcommand{\Tr}[1]{\mbox{Tr}\left(#1\right)}
\newcommand{\kb}{k_B}
\newcommand{\rad}{\mbox{ rad}}
\preprint{APS/123-QED}

\title{Friedel Oscillations in Graphene: Sublattice Asymmetry in Doping}

\author{J. A. Lawlor}
\affiliation{School of Physics, Trinity College Dublin, Dublin 2, Ireland}
\author{S. R. Power}
\email{spow@nanotech.dtu.dk}
\affiliation{Center for Nanostructured Graphene (CNG), DTU Nanotech, Department of Micro- and Nanotechnology,
Technical University of Denmark, DK-2800 Kongens Lyngby, Denmark}
\author{M. S. Ferreira}
\affiliation{School of Physics, Trinity College Dublin, Dublin 2, Ireland}
\affiliation{CRANN, Trinity College Dublin, Dublin 2, Ireland}

\date{\today}

\begin{abstract}
Symmetry breaking perturbations in an electronically conducting medium are known to produce Friedel oscillations (FOs) in various physical quantities of an otherwise pristine material.
 Here we show in a mathematically transparent fashion that FOs in graphene have a strong sublattice asymmetry.
 As a result, the presence of impurities and/or defects may impact the distinct graphene sublattices very differently.
 Furthermore, such an asymmetry can be used to explain the recent observations
 that Nitrogen atoms and dimers are not randomly distributed in graphene but prefer to occupy one of its two distinct sublattices.
 We argue that this feature is not exclusive of Nitrogen and that it can be seen with other substitutional dopants.    
\end{abstract}

                 
\maketitle

\section{Introduction}

\begin{figure}
\centering
\includegraphics[width = 0.45\textwidth]{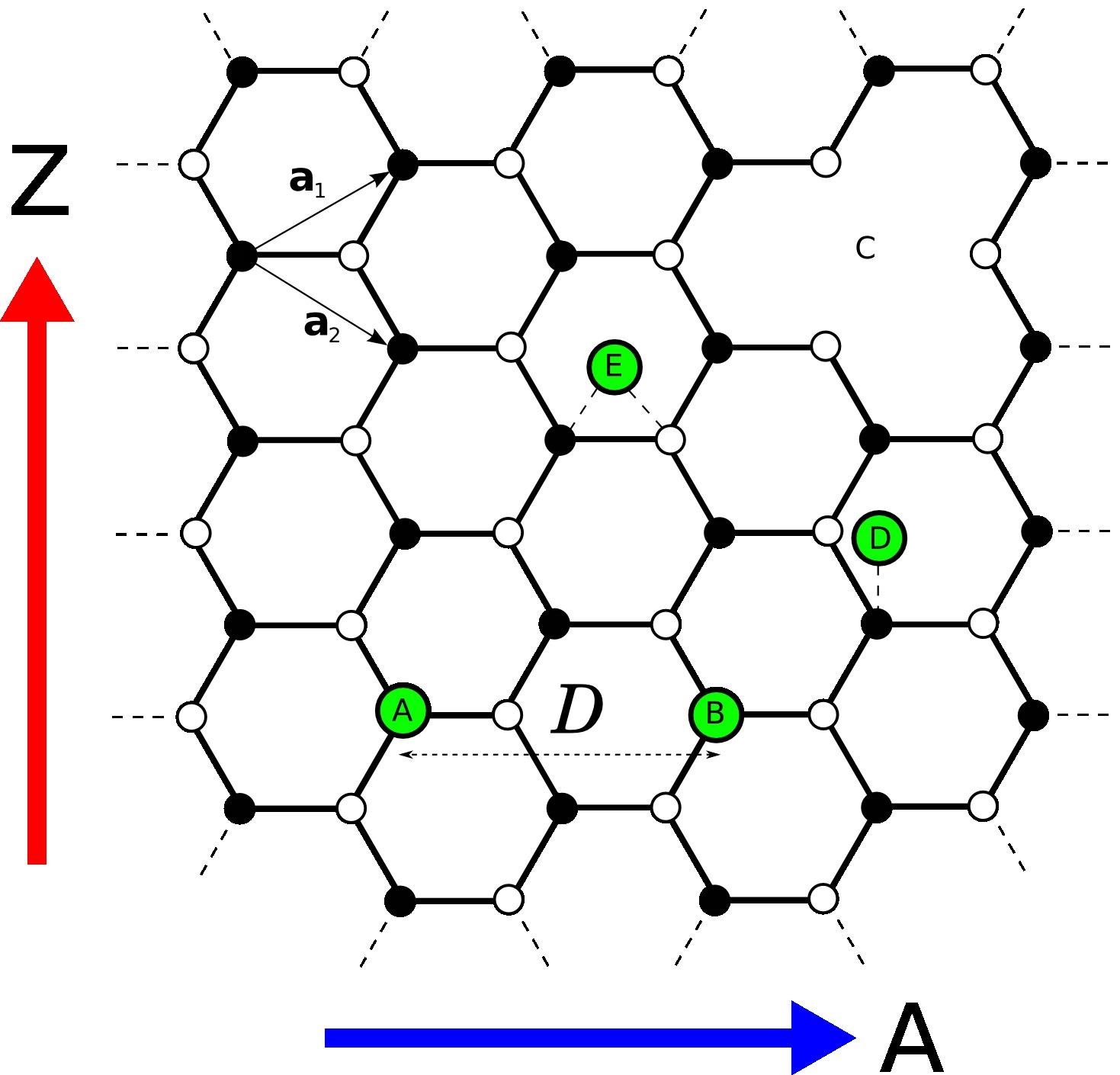}
\caption{Schematic of a graphene lattice with impurity bonding types, shown with the lattice vectors $\mathbf{a}_1$ and $\mathbf{a}_2$.
Graphene can be thought of as two interpenetrating triangular sublattices, black ($\bullet$) and white ($\circ$).
Shown are 2 substitutional impurities (A and B) separated by distance $D = 1$ in the armchair direction, and also examples of a vacancy (C), a top-adsorbed adatom (D) and a bridge-adsorbed adatom (E).}
\label{fig:graphene_schematic}
\end{figure}

Nanoscale characterization techniques are fundamentally based on the changes experienced by otherwise pristine materials in the presence
of symmetry-breaking impurities and defects. Such symmetry-breaking, for example in a Fermi gas, induces perturbations in the electronic environment of the gas through the scattering of its electrons \cite{friedel_original}.
These changes in the electronic scattering manifest as spatial oscillations, called Friedel oscillations (FOs), in quantities like the local density of states ($\rho$) and the carrier density ($n$), which radiate away from the location of the symmetry breaking perturbation and decay with the distance from the perturbation, $D$, with a rate linked directly to the dimensionality of the system and to some extent the resolution of the measurement.
Much attention has been focused recently on such symmetry breaking in graphene\cite{bacsiPRB, bacsi2, falkoFriedel, dutreiz_friedel_strain, tao_bergmann_fos, PhysRevB.75.241406, PhysRevLett.106.045504, CheianovEPJST}, a hexagonal lattice of sp-2 hybridised carbon with a wide range of unique properties\cite{graphene_original} . 
Fig. \ref{fig:graphene_schematic} shows a schematic of the lattice with several different kinds of impurities, where atomic sites in the two triangular interconnected triangular sublattices composing graphene are represented by black and white symbols.  
In graphene, the vanishing of the density of states at the Dirac Point affects the decay rate of the change in carrier density ($\Delta n$) from 
$D^{-2}$, expected in a 2-D system, to a faster $D^{-3}$ rate for ungated/undoped graphene and the oscillations disappear due to their commensurability with the lattice spacing\cite{bacsiPRB, PhysRevLett.106.045504, falkoFriedel}.

Previous studies examining the analytical behaviour of $\Delta n$ FOs in graphene have generally relied on a linearisation of the electronic bandstructure near the Dirac points and the introduction of a momentum cutoff \cite{falkoFriedel, bacsi2, bacsiPRB, PhysRevLett.106.045504}.  In the current work, we present an alternative framework which removes these assumptions and matches numerical results exactly in the long-distance limit and over large energy ranges, paving the way for applications to other electronic quantities and graphene-like materials. The methodology we use is similar to that used to describe the Ruderman-Kittel-Kasuya-Yosida (RKKY) interaction in graphene\cite{spaRKKYCalc, rkkySPACrystalsPub}, a coupling effect between magnetic impurities which has been studied extensively and, like FOs, is another manifestation of symmetry breaking, making this work a natural extension of those techniques. 

This methodology is applied to a range of commonly investigated impurity configurations, namely single and double substitutional impurities, vacancies, and realistic instances of the more commonly found top- and bridge-adsorbed atoms. The analytical expressions derived for the fluctuations in electron density for these impurities are corroborated with numerical calculations to confirm the predicted behaviour. We note how important features of the FO are dictated by the bipartite nature of the graphene lattice and furthermore, that different behaviours are observed for adsorbed impurities connecting symmetrically or asymmetrically to the two sublattices in graphene.
The framework is extended to consider similar oscillations which occur in the formation energy of two impurities introduced into the graphene lattice in close proximity to each other. Such FOs in formation energy are consistent with recent experimental findings of sublattice-asymmetric doping of nitrogen substitutional impurities in graphene\cite{nitrogen_original_science, nitrogen_second_experiment}. The ability to dope one sublattice of graphene preferentially opens many possibilities, including the opening of a bandgap \cite{nitrogen_doping_motivation}, and it is very interesting that such behaviour may be the manifestation of the strong sublattice dependence seen in FOs in graphene.

The paper is organised as follows. Section \ref{sec:methods} introduces the relevant mathematical methods required, in particular the Green functions for graphene and the perturbations associated with different impurity configurations. Section \ref{sec:fos_in_dn_dp} details the analytic approximations for the FOs in $\Delta n$ for the range of impurity type shown in Fig. \ref{fig:graphene_schematic} and compares their predictions to fully numerical calculations. 
As an application of our methodology Section \ref{sec:nitrogens} investigates the appearance of sublattice-asymmetry in nitrogen-doped graphene and presents a simple tight binding model for
the long-range sublattice ordering witnessed in recent experiments.

\section{Methods} 
\label{sec:methods}

\begin{table*}
  \centering
\scalebox{1.2}{
  \begin{tabular}{| c || p{2.2cm} | p{2.5cm} | p{2.2cm} | p{4cm} |}
\hline
      & & & & \\
            & \hspace{0.02cm} Substitutional & \hspace{0.4cm} Vacancy & \hspace{0.6cm} Top &  \hspace{1.2cm} Bridge \\[2ex] \hline
& 								  &  & & \\
    $\hat{V}$ &\hspace{0.45cm} $\lambda \ket{0} \bra{0}$  &  \hspace{0.0cm} $\lim_{\lambda \to \infty}\lambda \ket{0} \bra{0}$  &   \hspace{0.1cm} $  \ket{1} \tau \bra{a} + c.c. $         &  \hspace{0.2cm} $ \ket{1} \tau \bra{a} + \ket{2} \tau \bra{a} + c.c. $      \\[2ex] \hline
 & &  & & \\
    $\Delta G_{ii} $ &\hspace{0.45cm} $\frac{g_{i0}^2 \lambda}{1 - g_{00}\lambda}$ & \hspace{0.42cm} $\sim \frac{ g_{i0}^2}{-g_{00}} $     &  \hspace{0.1cm} $\frac{g_{i1}^2 |\tau|^2 g_{aa}} {1 - g_{aa} g_{11}|\tau|^2}$        & \hspace{0.1cm}  $\frac{(g_{i1}+g_{i2})^2 g_{aa} \tau^2}{1 - (g_{11}+g_{12}+g_{21}+g_{22}) g_{aa} \tau^2}$     \\[2ex] \hline 
  \end{tabular}}
  \caption{Perturbation operators ($\hat{V}$) and the corresponding fluctuation in the system GF at a general lattice site $i$ ($\Delta G_{ii}$).
 Our notation follows the rule that $a$ corresponds to an adsorbate atom site and the numerical indices specify the location of impurities - e.g. 0 for a substitutional impurity and 1 or 2 for the carbon host sites of adsorbates.}
  \label{tab:1}
\end{table*}

\subsection{Green Functions}

We begin by outlining the Green Functions (GFs) methods which play a central role in our approach as the Friedel oscillations in electron density and local density of states, $\Delta \rho$ and $\Delta n$ respectively, are directly obtainable using them.
The retarded single-body GF, $\mathcal{G}_{ij}$, between two cells $i$ and $j$ in the pristine lattice is calculated through diagonalisation of the nearest-neighbour tight-binding Hamiltonian using Bloch's Theorem and can be expressed as an integral over the Brillouin Zone \cite{spaRKKYCalc}

\begin{equation}
 \mathcal{G}_{ij} (E, \mathbf{r}) = \int \int_{B.Z.} \frac{d \mathbf{k}^2}{2 \pi^2} \frac{e^{i \mathbf{k . r}}}{E^2 - t^2 |f(\mathbf{k})|^2} \left[ \begin{array}{cc}
E & t f(\mathbf{k}) \\
t f^* (\mathbf{k}) & E\end{array} \right] ,
\label{equation:gf_integral}
\end{equation}
where the energy, E, includes an infinitesimal positive imaginary part,
 $t=-2.7$eV is the pristine graphene lattice hopping integral, $\mathbf{r} = m \mathbf{a}_1 + n \mathbf{a}_2$ (where $m,n \in \mathbb{Z}$) is the spatial separation of the unit cells $i$ and $j$ containing the relevant sites which is expressed in terms of the lattice vectors $\mathbf{a}_1$ and $\mathbf{a}_2$ shown in Fig. \ref{fig:graphene_schematic} and the variable $f(\mathbf{k}) = 1 + e^{- i \mathbf{k}.\mathbf{a}_1} +  e^{- i \mathbf{k}.\mathbf{a}_2}$.
 In our analytic work, we principally examine armchair direction separations, defined by $m = n$ and work in distance units of $\frac{m + n}{2}$.
 The matrix form of Eq. \eqref{equation:gf_integral} captures the inter-sublattice nature of the GF calculation between the two sites in the graphene unit cell,
 such that 
\[ \mathcal{G}_{ij}  (E, \mathbf{r})= \left[ \begin{array}{cc}
g_{ij}^{\bullet  \bullet}(E, \mathbf{r}) & g_{ij}^{\bullet  \circ}(E, \mathbf{r}) \\
g_{ij}^{\circ  \bullet}(E, \mathbf{r}) & g_{ij}^{\circ  \circ} (E, \mathbf{r})\end{array} \right] ,
\] 
where $g_{ij}^{s_1 , s_2}$ is the pristine lattice GF from the $s_1$ sublattice site in cell $i$ to the $s_2$ sublattice site in cell $j$. 
For conciseness we will omit the sublattice indices from hereon and use $g_{ij}$ to denote the pristine lattice GF between two sites on the same sublattice sites unless specified otherwise.

To aid numerical calculation one of the two integrals in Eq. (\ref{equation:gf_integral}) can be solved analytically, and
in some cases it is possible to approximate the second integral using the stationary phase approximation (SPA) \cite{spaRKKYCalc}.
These methods take advantage of the highly oscillatory nature of the integrand for large $\mathbf{r}$ and to approximate the integral at values near the stationary points of the phase, ignoring the highly oscillating parts which mostly cancel each other. 
The advantage of the SPA is that it is applicable across the entire energy band, with no intrinsic momentum or energy cut-offs,  although maximum achievable experimental gating (through ion gels) is limited to $E_F \sim 0.5|t|$ currently \cite{max_gating}. 
Applying the SPA assumption to $g_{ij}$ gives us a sum of terms in the form
\begin{equation}
\label{equation:spa}
 g_{ij} = \frac{ \mathcal{A}(E) e^{i Q(E) D}}{\sqrt{D}} \,,
\end{equation}
where the coefficients $\mathcal{A}(E)$ are dependent on the sublattice configuration of sites $i$ and $j$ and the direction between them. For two sites on the same sublattice separated in the armchair direction we find 
 \begin{equation}
\label{equation:ae}
 \mathcal{A}(E) = \frac{- 2 \sqrt{i E} }{\sqrt{\pi (E^2 + 3t^2) \sqrt{t^2 - E^2}}}   \,,
   \end{equation}
valid for $|E|<|t|$ where  $D = \frac{m+n}{2}$ is the separation between the sites $i$ and $j$ and $Q(E)$ is associated with the Fermi wavevector in the armchair direction. This
approximation works best for separations beyond a few lattice spacings in the armchair and zig-zag directions as other directions are not always analytically solvable \cite{spaRKKYCalc}. For an armchair separation of $D = 5$ the agreement between analytic and numerical calculations of $g_{ij}$ is excellent, with less than 1\% deviation over 95\% of the energy spectrum.

To calculate the fluctuations in properties of a system when an impurity is introduced, we will usually require the difference between the GFs describing the pristine ($\hat{g}$) and perturbed ($\hat{G}$) systems. This can be expressed, using the Dyson equation\cite{economou}, as
\begin{equation}
\label{equation:Dyson}
 \Delta \hat{G} = \hat{G} - \hat{g} = (\hat{I} - \hat{g}\hat{V})^{-1} \hat{g} - \hat{g} ,
\end{equation}
where $\hat{V}$ describes the potential applied to the pristine system to introduce the impurity. 
By finding suitable descriptions and parameterizations for different impurities in the graphene lattice we can find directly the change in the corresponding perturbed lattice GF $\Delta \hat{G}$, and also how the $\rho$ and $n$ are altered from the pristine system. Exact parameterization, however, is not
that important when considering the qualitative features of the phenomena we investigate here. More precise parameterization can be found, for example, by comparison to DFT calculations\cite{nitrogen_params}.

\subsection{Impurities}

In this section we present the types of impurities that will be considered and the respective perturbative potentials used to describe them.

Substitutional impurities in graphene, shown schematically by A and B in Fig \ref{fig:graphene_schematic}, occur when single carbon atoms are replaced with dopant atoms such as nitrogen. The simplest way to model them is  by introducing a perturbation $\hat{V}^{Subs.}$ (Table \ref{tab:1}) which alters
the onsite energy of the site in question by a quantity $\lambda$.
The onsite energy of the carbon sites neighbouring the impurity and their overlap integrals with the impurity site are presumed to be unchanged in this simple model. However they can be easily incorporated, as can additional orbitals beyond our single orbital approximation for the impurities, if a more accurate parameterization is required. 
The presence of a substitutional impurity at site $0$ induces a fluctuation of the diagonal matrix element GF, for example $\Delta G_{ii}$ at site $i$, 
which can be found through applying $\hat{V}^{Subs.}$ to Eq. \eqref{equation:Dyson} the result of which is shown in Table \ref{tab:1}.

Vacancy defects, formed by the removal of a carbon atom from the graphene lattice (Fig. \ref{fig:graphene_schematic} (C)), can be considered as a phantom substitutional atom in the limit where the onsite energy,$\lambda \rightarrow \infty$. In a physical context this is effectively excluding the state to the electrons in the lattice. Experimentally, vacancies can be induced by ion bombardement \cite{ion_bombard}.
An approximation of the corresponding $\Delta G_{ii}^{Subs.}$, using the limit $\lambda \to \infty$, is shown in Table \ref{tab:1}. 
Alternatively this can be modelled by removing the hopping between the site and the nearest neighbours. 

Adsorbates, which bind to atoms in the graphene lattice, are characterised by an atom with onsite energy $\epsilon_a$ which is initially disconnected from the graphene sheet with an onsite GF $g_{aa} = 1/(E - \epsilon_a)$. 
A top-adsorbed adatom (Fig \ref{fig:graphene_schematic} (D)) is connected to the lattice by the perturbation $\hat{V}^{Top}$ in Table \ref{tab:1} 
which connects the adatom and site $1$ in the graphene lattice, where $\tau$ is the overlap integral between the adatom and the host lattice site. 
Similarly, a bridge-adsorbed adatom (Fig. \ref{fig:graphene_schematic} (E)) is attached to two carbon host sites in the lattice through an extra term in $\hat{V}$. The onsite GF fluctuations $\Delta G_{ii}^{Top}$ and $\Delta G_{ii}^{Bridge}$ are again shown in Table \ref{tab:1}.

\subsection{Charge Density Perturbations}
The effect of a general perturbation $\hat{V}$ on the charge density on site $i$ ($\Delta n_i$)  is calculated through an integration of the change in LDOS, $\Delta \rho_i$,
\begin{equation}
 \label{equation:delta_n}  \Delta n_i = \int_{-\infty}^{\infty} dE f(E) \Delta \rho_i \,,
\end{equation}
where $f(E)$ is the Fermi Function and $\Delta \rho_i = - \frac{2}{\pi} \text{Im} \Delta G_{ii}$ relates the change in local density of states 
to the perturbed diagonal GF $\Delta G_{ii}$. 
Physically, $n_{i}$ is the number of states below the Fermi energy ($E_F$) that are filled by electrons on site $i$, where $\rho_i$ describes the energy distribution of these states at the site. When calculating $\Delta n$ numerically the integral is evaluated along the imaginary energy axis to avoid discontinuties along the real axis. This transformation is done through forming a contour in the upper half plane, which contains no poles, and evaluating the contour integral via Cauchy's Theorem.

\subsection{Changes in Total System Energy} \label{sec:lloyd}
Green Functions methods can also be used to quantify other phenomena associated with the lattice, for example the change in the total system energy
due to the bonding of impurities follows from a sum rule and is given explicitly by the Lloyd Formula \cite{lloyd_formula}
\begin{equation}
 \label{equation:lloyds}
 \Delta E = \frac{2}{\pi} \text{Im}  \int_{-\infty}^{\infty}  dE f(E) \ln det ( \hat{I} - \hat{g}\hat{V} ).  
\end{equation}
This quantity is directly related to $\hat{V}$ and is useful for finding energetically favourable impurity positions
and can be used to investigate the dispersion and clustering of impurities\cite{impurity_segregation, impurity_seg_2}.
For multiple impurities the expression will contain interference terms which result in changes from the single impurity case and can reveal favoured configurations in the lattice.

\section{Friedel Oscillations in Charge Density and LDOS} 
\label{sec:fos_in_dn_dp}

\subsection{Weak Substitutional Impurities}  
\label{sec:one_subs_imp}

\begin{figure}[ht!]
\centering
\includegraphics[width = 0.45\textwidth]{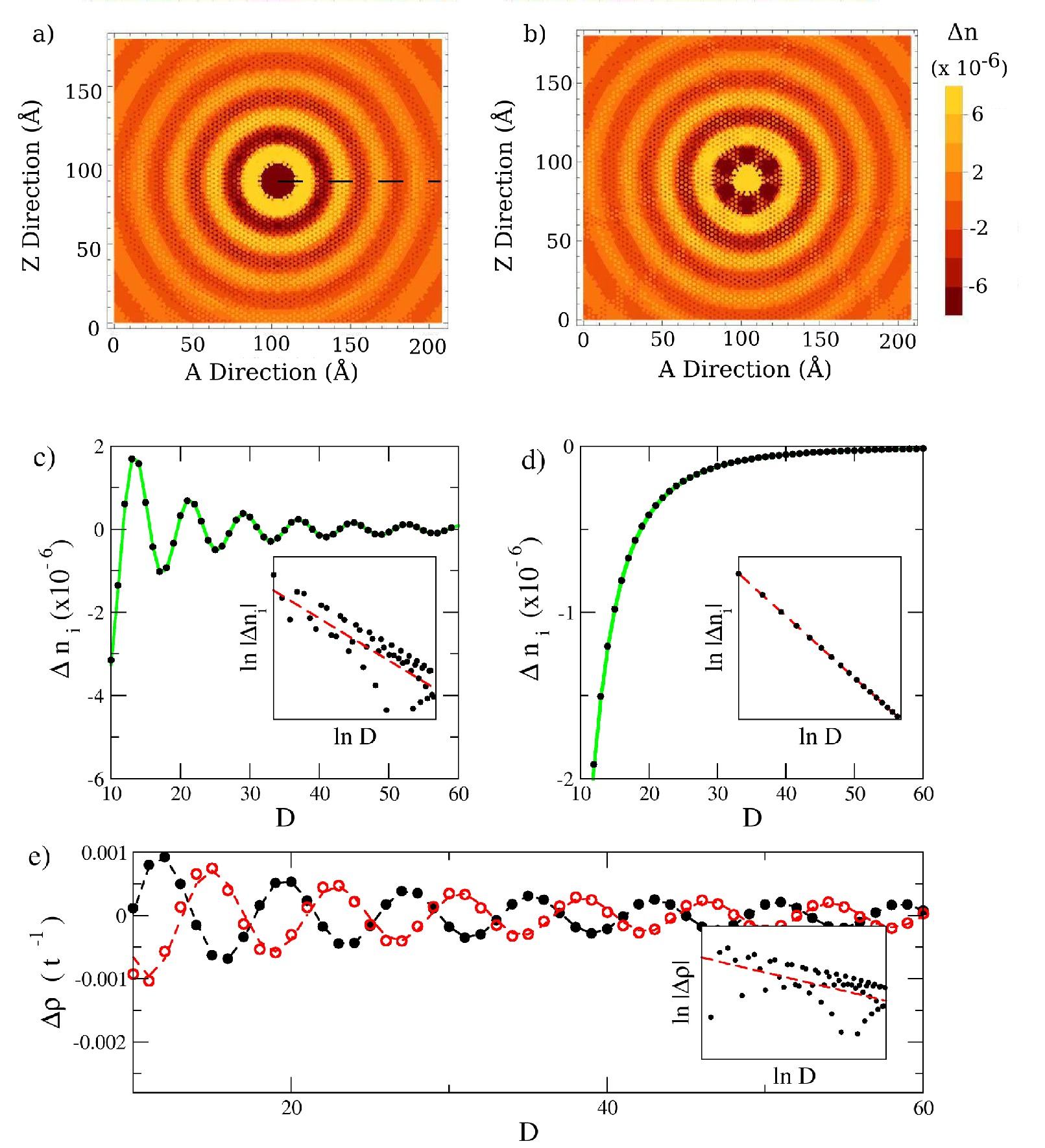}
\caption{The FO introduced by a substitutional impurity of strength  $\lambda = |t|$. 
Numerical Contour Plots of FOs in $\Delta n$ on the black (a) and white (b) sublattices at $E_F = 0.2|t|$ with the impurity located at the center,
calculated using Eq. (\ref{equation:delta_n}). Plot (c) shows a cross-section of numerical (black circles) and SPA (green line) 
calculations of $\Delta n$ on the black sublattice along the dashed line shown in panel (a) for $E_F = 0.2|t|$. The inset shows a log-log plot of this
data with a regression line (red, dashed) corresponding to a decay of $D^{-2}$. (d) shows $\Delta n$ at the same locations when $E_F = 0$ where we 
see an absence of oscillations and a quicker decay, shown by the  $D^{-3}$ regression line in the inset. Panel (e)
is a cross-section of $\Delta \rho_i$ for the same impurity and locations, with numerical results for the black (white) sublattice given by the black, solid (red, hollow) symbols. Analytic SPA results are given by the corresponding dashed lines, and the inset shows a log-log plot of the data with a red dashed line corresponding to a decay of $D^{-1}$.}
 \label{fig:contour_plot}
\end{figure}

To begin, we consider the charge density variations $\Delta n$ at all lattice sites surrounding a substitutional impurity of 
strength $\lambda = |t|$, situated on a site in the black sublattice. A numerical evaluation using Eq. (\ref{equation:delta_n}) and  
$\Delta G_{ii}^{Subs.}$ from Table \ref{tab:1} yields the contour plots in Fig \ref{fig:contour_plot} (a) and (b), where 
we see FOs in the charge density radiating away from the central impurity on both sublattices with a wavelength determined by $E_F$.
There is clear sublattice asymmetry in $\Delta n$ between the black (a) and white (b) sublattices with $\Delta n$ swapping signs between the sites in the same unit cell. This signature is important when considering multiple impurities, which we will discuss in sections \ref{sec:mult} and \ref{sec:nitrogens}. 
It is possible to approximate $\Delta \rho$ and $\Delta n$ along the armchair direction (dashed line in Fig. \ref{fig:contour_plot} (a)),
by applying the SPA approach and the Born Approximation, which is valid for weak scatterers of strength $\lambda \lesssim |t|$, to $\Delta G_{ii}^{Subs.}$,  
resulting in
 \begin{equation}  
 \Delta \rho_i \approx \frac{-2}{\pi} \text{Im} g_{i0}^2  
 \label{eq:ldos_born}
 \end{equation} 
\begin{equation}
\label{equation:subs_fo}
\Delta n_i^{Subs.} \approx -\lambda \frac{2}{\pi}  \text{Im}  \int_{-\infty}^{\infty} dE f(E) g_{i0}^2  \,,
\end{equation}
where $g_{i0}^2$ can then be expressed using Eq. (\ref{equation:spa}) as $\frac{A^2 e^{2 i Q(E) D}}{D}$ and we can solve the integral 
via contour integration in the upper-half of the complex energy plane, the poles in the integrand being given by the Matsubara frequencies. 
Taking the limit of zero temperature gives the sum
\begin{equation}
\label{equation:spa_fo_one} 
\Delta n_i^{Subs.} \sim \lambda  \frac{2}{\pi}\text{Im}  \sum_{\ell=0}^{\infty} \frac{ \gamma_\ell (E)  } { D^{\ell+2} } e^{2 i Q^{(0)} D} \,.
\end{equation}
The sum coefficients $ \gamma_\ell (E)$ are related to the SPA coefficients and defined as
\begin{equation}
\label{equation:gamma} \gamma_\ell (E) = \frac{(-1)^\ell \mathcal{B}^{(\ell)} } {(2 Q^{(1)})^{\ell+1}} ,
\end{equation}

where $\mathcal{B} = \mathcal{A}^2$ and $X^{(\ell)}$ denotes the $\ell^{th}$ derivative of the function $X$ with respect to energy. 
$\mathcal{B}^{(\ell)}$, $Q^{(0)}$ and $Q^{(1)}$ are evaluated at the Fermi energy $E_F$. The first order term of $\Delta n$ thus decays as
$D^{-2}$ with an oscillation period determined by $Q^0$, and thus $k_F$ (Fig. \ref{fig:contour_plot} (c)).
At the Dirac point we find that $Q^0 \to \pi$ in the phase factor which causes the sign-changing oscillations to become commensurate with the lattice spacing and seemingly disappear for all terms. Additionally, the energy dependent term $B^0 \to 0$ and so the leading term of the series $\gamma_0 = 0$. Taking the next leading term of the series in Eq. \ref{equation:spa_fo_one} ($\ell = 1$) gives $\Delta n_i$ decaying as $D^{-3}$ and a comparison between the SPA approximation and the numerical result is shown in Fig. \ref{fig:contour_plot} (d).
We also note that  $\Delta \rho$ decays as $D^{-1}$ as shown in Fig. \ref{fig:contour_plot} (e) away from the Dirac Point, which can be inferred directly from Eq. \eqref{eq:ldos_born}. The sublattice asymmetry is quite clear in the cross-section of $\Delta \rho$, which is shown for sites on the same (black) and opposite (red) sublattice as the impurity. The opposing sign of the oscillations is a signature of FOs in $\Delta \rho$, $\Delta n$, $\Delta E$ 
and also of the RKKY interaction\cite{rkkySPACrystalsPub}. 
These results agree with previous work \cite{bacsiPRB, falkoFriedel}, but due to our approximation method requiring large $D$ we are limited to the long-range behaviour in the ranges of 5+ unit cells and so miss short wavelength features which are present in the region immediately surrounding an impurity which have been investigated in more depth by Bacsi and Virosztek \cite{bacsiPRB, bacsi2}. In addition to what is shown in Fig. \ref{fig:contour_plot} we found excellent agreement between numerical  and analytic calculations for all energy and sublattice configurations.

\subsection{Multiple Weak Substitutional Impurities} \label{sec:mult}
\begin{figure}[ht!]
\includegraphics[width = 0.45\textwidth]{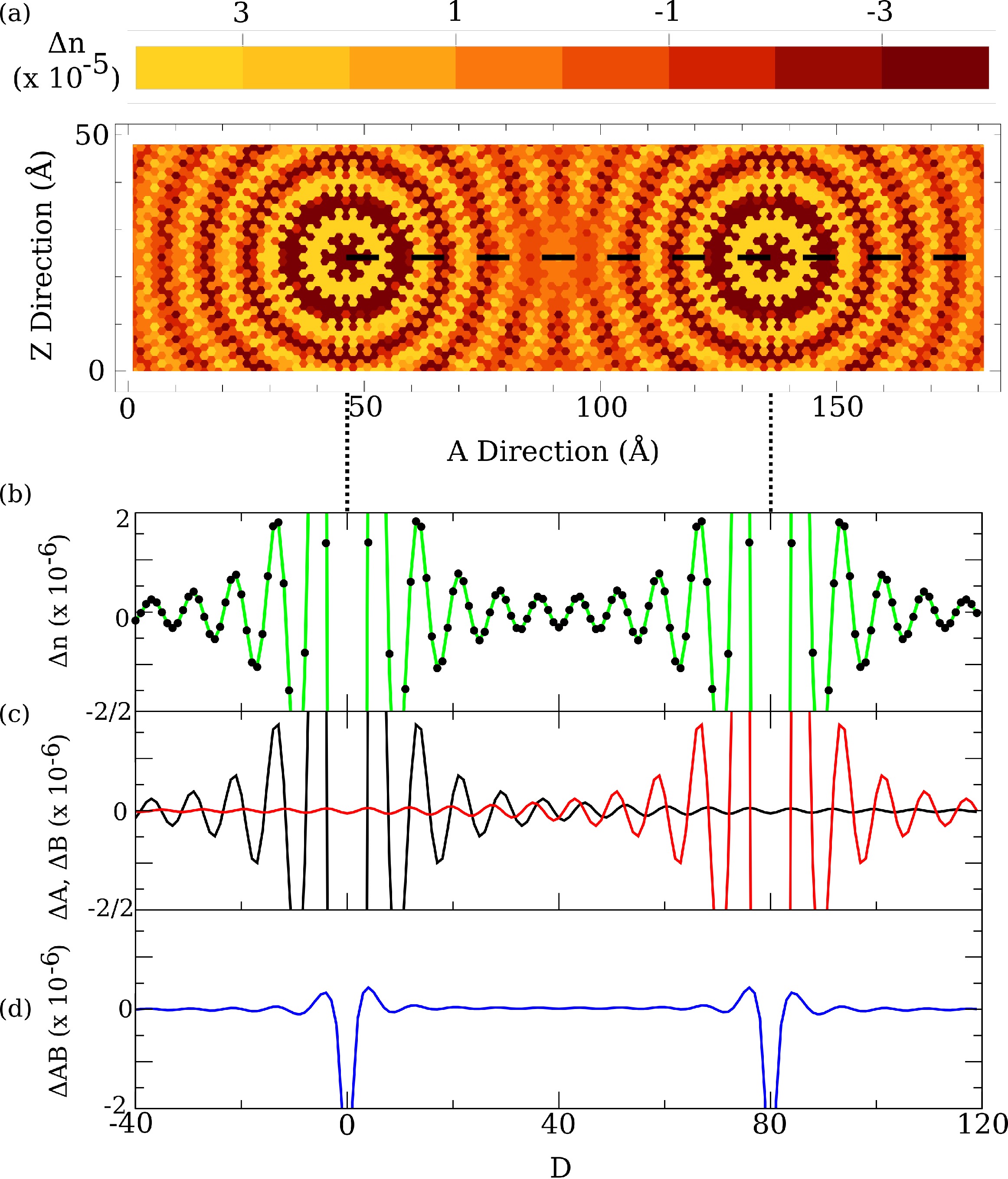}
\caption{(a): Numerical simulation of $\Delta n$ FOs on the black sublattice for two weak scatterers $\lambda = 0.1 t$ separated by 40 unit cells in the armchair direction  at $E_F = 0.2|t|$. As the chosen Fermi Energy lies in the linear dispersion regime, the interference pattern that arises is similar to that seen in classical waves. (b) Cross section of $\Delta n$ along the dashed line in the top contour plot of analytic (green) and numerical (black symbols) calculations. (c) and (d): Contributions to $\Delta n$ from the terms $\Delta_A$ (black), $\Delta_B$ (red) and $\Delta_{AB}$ (blue) per Eq. (\ref{equation:dn_multiple}).}
\label{fig:two_impurities}
\end{figure}
When considering two or more substitutional impurities we can extend the matrix $\hat{V}^{Subs.}$ (Table \ref{tab:1}) to include additional sites by addition of extra perturbations at the corresponding locations, for example for two identical impurites at arbitrary sites $A$ and $B$ we have $\hat{V} = \lambda \ket{A}  \bra{A} + \lambda \ket{B} \bra{B}$.
Fig. \ref{fig:two_impurities} (a) shows a contour plot of $\Delta n$ on the black sublattice for two such impurities spaced by $D = 80$ in the armchair direction, where the SPA can be used to approximate $\Delta n$ along the dashed line. Generally, if $\lambda \lesssim t$ then we find through Eq. (\ref{equation:delta_n}) and the Born Approximation that to first order $\Delta n$ has the form
\begin{equation}
 \label{equation:dn_multiple} \Delta n_{i}^{Mult.} \approx -\frac{2}{\pi} \text{Im}  \int dE f(E) ( \Delta_{A} + \Delta_{B}  + \Delta_{AB} ) ,
\end{equation}
where
\[ \Delta_A = g_{iA} \lambda g_{Ai},\]
\[ \Delta_B = g_{iB} \lambda g_{Bi},\]
and
\[ \Delta_{AB} = 2 g_{iA} g_{AB} g_{Bi} - g_{iA} g_{BB} g_{Ai} - g_{iB} g_{AA} g_{Bi} .\]

$\Delta_A$ and $\Delta_B$ arise from the effects of the individual isolated impurities and the extra term  $\Delta_{AB}$ describes the interference effect. 
The approach for the single impurity can be adapted and applied to all three terms, where the integration of $\Delta_A$ and $\Delta_B$ will
be identical to the single impurity case and the interference term $\Delta_{AB}$ can be approximated well if $A$ and $B$ are at least a few unit cells apart.
Clearly $\Delta_{AB}$ is dependent on $\lambda^2$, but also on the separation of A and B. Thus this term decays rapidly when the scatterers are
weak and/or separated by several unit cells. Consider a cross-section of the charge density fluctuations indicated by the black dashed line in Fig. \ref{fig:two_impurities} a). Applying the SPA derived in the previous section we can match the interference pattern seen, as shown in Fig. \ref{fig:two_impurities} b), achieving an excellent match between analytic and numerical methods.
By breaking apart the SPA terms we find that the dominant contribution is from the isolated impurities (given by $\Delta A$ (black curve) and $\Delta B$ (red curve) in Fig. \ref{fig:two_impurities} c)) with a very small contribution from the interference term $\Delta AB$ regardless of chosen energy (Fig. \ref{fig:two_impurities} d)).

\begin{figure}[ht!]
\includegraphics[width = 0.45\textwidth]{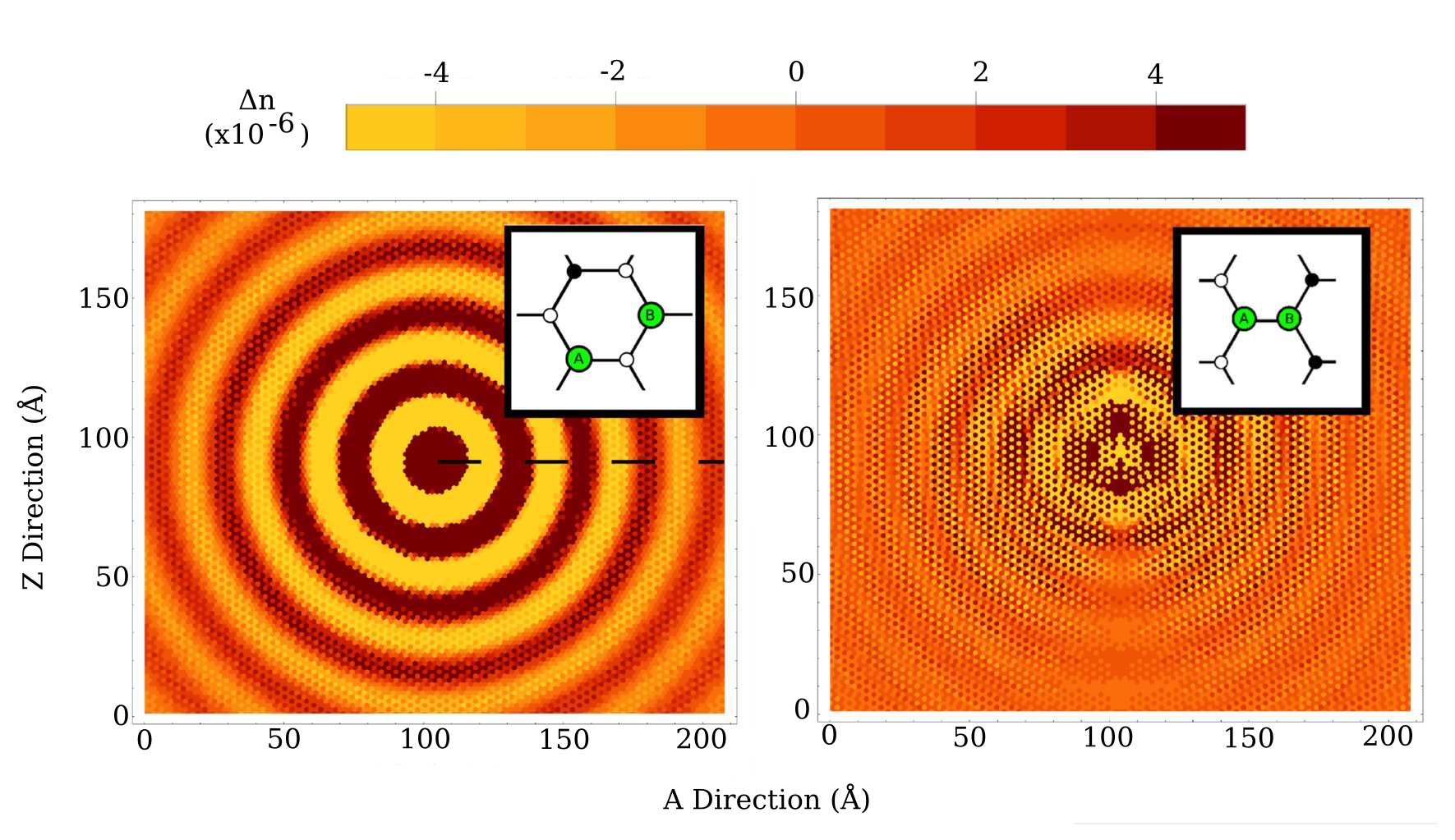}
\caption{ Black sublattice plots for quasi-nearest neighbour impurities (left) and impurities sharing the same unit cell (right) as demonstrated in 
the insets for $\lambda = t$ and $E_F = 0.2|t|$.
 The pairwise interaction and opposite sign of each impurity's individual contribution leads to a complex $\Delta n$ pattern in the second case.}
\label{fig:neighbours}
\end{figure}

Whilst the SPA works well when $D$ is suitably large and $\lambda$ small, the simple approximation breaks down when the impurities are moved to within a couple of unit cells as the contribution of the interference terms becomes more important, especially for impurities on opposite sublattices. In addition to the interference term $\Delta_{AB}$, the region of significant overlap between the $\Delta_A$ and $\Delta_B$ increases and the FO patterns observed become more complex. In Fig. \ref{fig:neighbours} we examine the numerical contour plot on the black sublattices of $\Delta n$ for two such configurations, namely two substitutional atoms of the type considered in Fig. \ref{fig:contour_plot} which are either next-nearest neighbours residing on the same sublattice (left panel) or nearest neighbours on opposite sublattices (right panel). These configrations are shown schematically in the insets. For the first case, we note the strongly sublattice dependent behaviour noted in Section \ref{sec:one_subs_imp} is 
still present, whereas in the nearest-neighbour impurity case it has been mostly washed out due to a superposition of the features observed in panels a) and b) of Fig. \ref{fig:contour_plot}. 
The strikingly different interference patterns present clear signatures for the two cases and this general qualitative difference in FOs may make impurity configurations easier to distinguish. The importance of this cross sublattice effect is apparent when considering FOs in other quantities, and will be discussed later in the context of energetically favourable doping configurations for multiple nitrogen substitutional impurities in graphene.

\subsection{Vacancies and Strong Scatterers} 
\label{sec:vacancies}
Taking the limit $\lambda \to \infty$ for a substitutional impurity, as shown in Table \ref{tab:1}, corresponds to placing a vacancy in the lattice and yields
\begin{equation}
\label{equation:vacancies} \Delta n_i^{Vac.} \approx  \frac{-2}{\pi}  \text{Im}  \int_{-\infty}^{\infty} dE \frac{f(E) g_{i0}^2}{-g_{ii}}.
\end{equation}
We note that the change in charge density on the impurity site, given by $g_{i0} = g_{00}$ in Eq. (\ref{equation:vacancies}), becomes $\Delta n_0 = -1$
which corresponds to a complete depletion of electrons on this site. 
The pristine onsite $g_{ii}$ can be approximated very well for energies in the linear regime as \cite{katsnelsonbook}
\begin{equation}
 \label{equation:gii_approx} g_{ii} \approx  \frac{2}{\sqrt{3} \pi t^2} E \ln{\frac{|E|}{ 3 t}}- i \frac{|E|}{\sqrt{3} t^2},
\end{equation}
and this approximation works at energies up to approximately $E_F \sim |t|/2$.
Eq. (\ref{equation:vacancies}) can be solved in a similar fashion to that of the single substitutional impurity by observing that $g_{ii}$ is a function of energy only and absorbing it into the usual $\gamma_\ell$ term in Eq. (\ref{equation:spa_fo_one}) and Eq. (\ref{equation:gamma}), then following through with the usual derivation. There is no pole in the upper half plane for this integrand, so the evaluation method remains unchanged from the weak impurity case.

\begin{figure}[ht!]
\includegraphics[width = 0.45\textwidth]{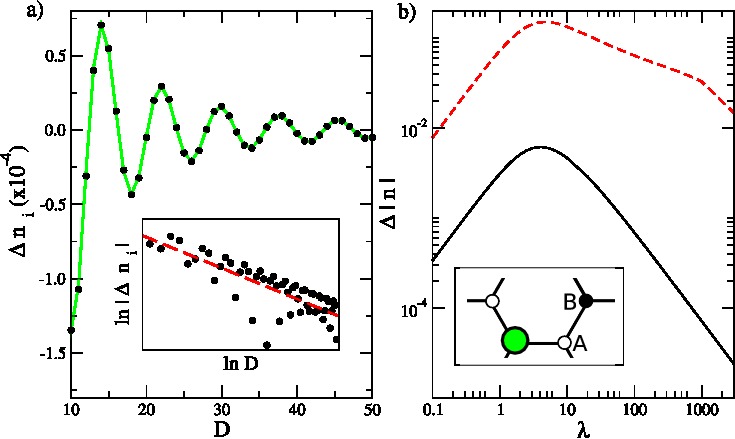}
\caption{a) Cross-section of FOs in $\Delta n$ at $E_F = 0.2|t|$ on the black sublattice in the armchair direction due to a vacancy at site 0
for both numerical (black) and analytic (green) calculations with a logarithmic plot inset showing $D^{-2}$ decay indicated by the red dashed line.
b) Log-log plot of $| \Delta n_1 |$ vs $\lambda$ at $E_F = 0$ on $A$ (black line) and $B$ (red line) sites shown in the inset schematic.
 For $\lambda \gg 1$ and $\lambda \ll  1$, 
the amplitude of $\Delta n_1$ becomes very small with a maximum amplitude at $\lambda \approx 5t$, with a positive sign on $B$ and a negative sign on $A$.}
\label{fig:numerical_vs_spa}
\end{figure}

Vacancies can be calculated numerically by either inducing a very large $\lambda$ value on a site, or by disconnecting the site from its neighbours.
A comparison of this numerical calculation and the SPA approximation of $\Delta n_i^{Vacancy}$ (Table \ref{tab:1}) on sites within the same sublattice as the impurity is shown for a finite $E_F$ in Fig. \ref{fig:numerical_vs_spa} (a), and excellent agreement is seen between the two. The general features are similar to those noted for weaker impurities, namely sign-changing oscillations and a $D^{-2}$ decay. 

Interestingly, when we set $E_F = 0$ we find a complete absence of FOs with $\Delta n = 0$ at all sites, corresponding to no change in $n$ from the pristine state.
 This behaviour is noted whether we examine sites on the same or on the opposite sublattice to the impurity,
 and is in stark contrast to the case of weaker scatterers where a non-oscillatory $D^{-3}$ behaviour was noted.
To try to understand this unusual behaviour, we examine the imaginary-axis integral that we need to solve to find $\Delta n$, which is of the form
\begin{equation}
 \label{equation:dn_vacancy_complex} \Delta n_i^{Vac.} \sim Re \int_{\eta}^{\infty} dx \frac{g_{i0}^2 (E_F + ix)}{ g_{00} (E_F + ix)} \,.
 \end{equation}
We will see that it is the behaviour of the individual GFs along the imaginary axis which determines the presence or absence of the carrier density oscillations. 
At $E_F = 0$ both $g_{00}$ and $g_{i0}$, if $i$ is on the black sublattice, become entirely imaginary along the integration range. Similarly if $i$ is on the
white sublattice, $g_{i0}$ is entirely real, ensuring that on both sublattices the integrand becomes entirely imaginary over the integration range and
$\Delta n$ vanishes since it depends on the real part of the integrand only. 

In the limit of large values of $\lambda$ the perturbations described by Eq.(\ref{equation:dn_vacancy_complex}) will vanish.
In Fig. \ref{fig:numerical_vs_spa} (b), we examine the charge density fluctuations on the site nearest the impurity on both the same (black curve, A) and opposite (red, dashed curve, B) sublattice as $\lambda$ is increased at $E_F = 0$. These sites will have the largest change in carrier density due to their proximity to the impurity. We note that $\lambda \approx 4t$ ($\lambda \approx 10t$) causes the largest amplitude in $\Delta n$ on the black (white) sublattice, with the amplitude of $\Delta n$ decreasing as $\lambda$ increases further . 
It should be emphasised that this disappearance of $\Delta n$ in the $\lambda \to \infty$ limit occurs only at $E_F = 0$, whereas other energies will show the familiar $D^{-2}$ oscillatory pattern.
\par
It is straightforward to evaluate $\Delta \rho_i$ from the $\Delta G_{ii}^{Vacancy}$ approximation in Table \ref{tab:1}, 
where there is a clear asymmetry between the black and white sublattices at the Dirac Point, as $g_{i0}^{\bullet \bullet} \to 0$ (with $\bullet$ being the minority sublattice - a vacancy removes a carbon atom from the sublattice) and $g_{i0}^{\bullet \circ} \to + i \infty$ ($\circ$ is the majority sublattice as it has more carbon atoms).
 This results in a zero density of states, similar to undoped graphene, on the minority sublattice and leads to divergencies in $\Delta \rho$ on the majority sublattice,
 which correspond to the widely predicted midgap resonance states seen in many previous works\cite{wehlingVacancyLDOS, vacancy_resonance1, vacancy_resonance2}.
This is an explanation of the phenomenon of vanishing $\Delta n$ for the vacancy case as mentioned previously in this section,
 where the vacancy introduces a resonance peak at $E = 0$ in the LDOS of the
majority sublattice sites with the bound state outside the band being removed to infinity.
 As the total number of states must be preserved such that $\frac{-2}{\pi} Im \int_{-\infty}^{\infty} dE \rho = 2$ at all sites the deformation of the
LDOS by the vacancy to form the peak at $E = 0$ ensures that when $E_F = 0$ the filled electron and hole states are symmetrical as in the pristine case leading to 
$\Delta n = 0$ for these cases. On the minority sublattice the LDOS is symmetrical about $E = 0$ but without the resonance so again electron-hole symmetry is preserved.
 Once more the sublattice nature of graphene introduces significant asymmetries in the features of FOs surrounding a defect.

\subsection{Adsorbates}

An analytic expression for $\Delta n_i^{Top}$, the FO induced by a top-adsorbed impurity (as shown in Fig \ref{fig:graphene_schematic} (d)), 
can be found using similar methods to that of a single substitutional atom, since $\Delta G_{ii}^{Top}$ in Table \ref{tab:1} is
analytic in the upper half plane and there are no additional poles other than those at the Matsubara Frequencies. 
However, an approximation of $g_{11}$ as in Eq. (\ref{equation:gii_approx}) is required as this GF is beyond the scope of the SPA. 
This approximation restricts our results to the linear regime.

\begin{figure}[t]

\includegraphics[width = 0.45\textwidth]{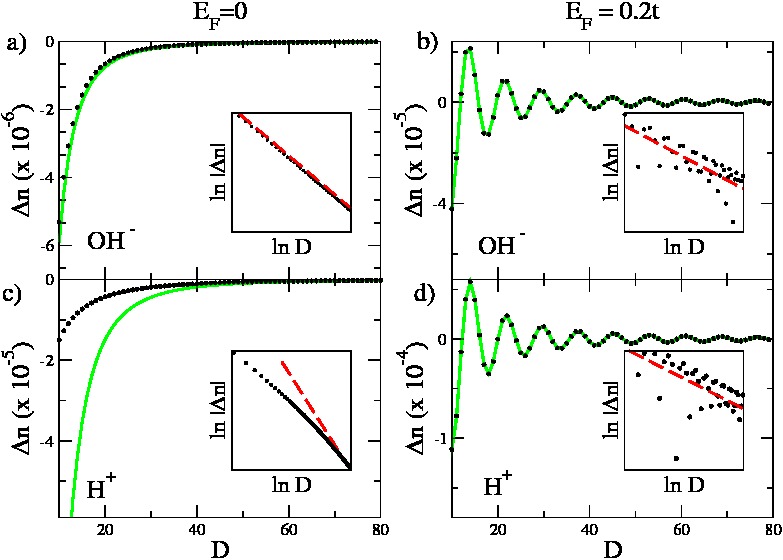}

\includegraphics[width = 0.45\textwidth]{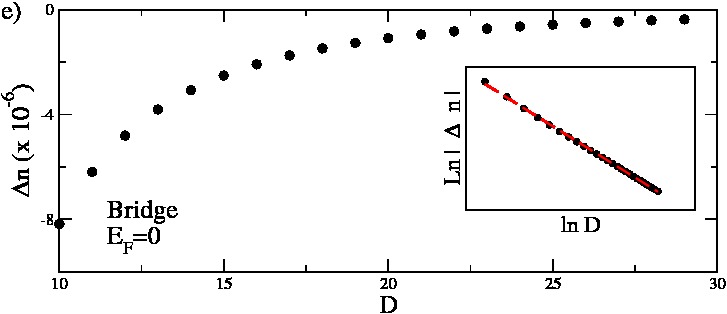}

\caption{FOs in $\Delta n$ due to top adsorbed $OH^-$ (top row) and  $H^+$ (middle row) and to bridge-adsorbed carbon (bottom panel). The top two cases show both $E_F = 0$ (left panels) and $E_F = 0.2|t|$ (right panels).
 Numerical data (black) is compared with the analytic expression (green), with corresponding 
log-log plots as insets. The red lines in the insets correspond to $D^{-2}$ fits for both $E_F = 0.2|t|$ plots (panels (b) and (d)) and $D^{-3}$ for the other cases. 
In panel c), we note that due to the parameterization of the $H^+$, the SPA is a poor match to the numerics at $E_F = 0$ away from the asymptotic limit.
The charge density perturbations for a bridge-adsorbed carbon adatom, shown here as a numerical calculation (black) in the bottom panel,  do not vanish at the Dirac Point, unlike a top adsorbed carbon, due to cross sublattice interferences.}
\label{fig:both_adsorbates}
\end{figure}

We consider two realistic cases of hydroxyl ($OH^-$) and hydrogen ($H^+$) adsorbates, using the parameters  $\epsilon_{OH-} = -2.9t \,, \,\tau_{OH-} = 2.3 t$ for hydroxyl\cite{oh_group_parameters} and $\epsilon_{H+} = -t/16 \,, \, \tau_{H+} = 2t$ for hydrogen\cite{hydrogen_parameters}.
The results for these impurities for $E_F = 0$ and $E_F = 0.2|t|$ cases are presented in Fig. \ref{fig:both_adsorbates}. 
We note an excellent match between the numerical (black circles) and analytic (green lines) approaches for all cases with the exception of Hydrogen at $E_F = 0$. 
The parameterization of hydrogen leads to divergencies at $E_F = 0$ which requires us to look much further away to see agreement (approx. 100 unit cells). 
The proximity of the hydrogen onsite term to the Dirac point produces a resonance condition at $E_F = 0$ and we see a clear change in the decay rate in the region close to the adsorbate in Fig. \ref{fig:both_adsorbates},  with an eventual $D^{-3}$ rate far from the defect. This has been studied recently by Mkhitaryan and Mishchenko \cite{resonance_decay_change}.
In the case $E_F \neq 0$ a decay rate of $D^{-2}$ is found for both adsorbates, with $D^{-3}$ at the
Dirac Point, with proviso, matching the behaviour of a substitutional impurity which could be expected due to the possibility of modelling the effect
of the adatom through a self-energy term replicating a substitutional atom\cite{economou}.

We note that altering the onsite energy to $\epsilon_a = E_F = 0$ forces a resonance condition and confirms previous findings
where the adatom behaviour can be similar to that of a vacancy \cite{hydrogen_parameters} and the $\Delta n$ FOs disappear completely, 
which we may expect, for example, in the case of a top adsorbed carbon. 
However, carbon prefers a bridge adsorbed configuration \cite{carbon_bridge_config} and due to this bonding type the interference effects from the two host sites, which are on opposite sublattices,  lead to finite charge density perturbations at the Dirac Point.
The presence of FOs in $\Delta n$ for a bridge adsorbed carbon at $E_F = 0$ can be seen by considering an arbitrary site $i$ and the
corresponding $\Delta G_{ii}^{Bridge}$ (Table \ref{tab:1}).
In section \ref{sec:vacancies} we noted that, when considering imaginary axis integration of off-diagonal GFs at $E_F = 0$, that $g_{ab}$ is either entirely real  (for opposite sublattice propagators) or entirely imaginary (for same sublattice propagators). However from the form of $\Delta G_{ii}^{Bridge}$ it is clear that there will exist both real and imaginary terms in the integrand for $\Delta n_i$, which was not the case for top-adsorbed or vacancy impurities. 
This cross sublattice interference is key to the non-vanishing $\Delta n$ FOs for all bridge adsorbed atoms, regardless of parameterization,
and in Fig. \ref{fig:both_adsorbates}(e) we show a numerical plot of the familiar $D^{-3}$ decay in $\Delta n$ at the Dirac Point. The usual $D^{-2}$ oscillations are recovered with doping.

\section{Sublattice Asymmetry in Nitrogen Doped Graphene} \label{sec:nitrogens}
Recent experimental works involving substitutional nitrogen dopants in graphene have reported a distinct sublattice asymmetry in their distribution, where the impurities are discovered to preferentially occupy one of the two sublattices, instead of being randomly distributed between them. This effect is noted at both long and short ranges\cite{nitrogen_original_science} and is corroborated with DFT results\cite{nitrogen_second_experiment}.  A distinct and controllable sublattice asymmetry in doped graphene presents many interesting possibilities, among them the possibility of inducing a band gap by controlling the dopant concentration - an important step in the development of graphene-based field-effect transistors\cite{nitrogen_doping_motivation}. 
As remarked in the Introduction, Green Functions methods can be extended beyond FOs in $\Delta n$ and $\Delta \rho$ to include other quantities such as the change in total system energy ($\Delta E$), due to a perturbation $\hat{V}$, using Eq. (\ref{equation:lloyds}). 
The calculation of $\Delta E$ allows the investigation of favourable impurity configurations, and we will now apply this method to study substitutional nitrogen impurities in graphene within a simple tight-binding model where the nitrogen impurities are characterised\cite{nitrogen_params} by $\lambda_N = -10 \text{eV}$.

\subsection{Total Energy Change}
We will first consider the interaction between two identical substitutional impurities at sites A and B with onsite energies $\lambda$, as in the multiple scattering case discussed previously,  so that the determinant in Eq. (\ref{equation:lloyds}) becomes $det ( \hat{I} - \hat{g}\hat{V} ) = (1 - g_{AA} \lambda)(1 - g_{BB} \lambda) - g_{AB}^2 \lambda^2$.
It is possible then to identify two separate contributions to $\Delta E$. The first one is associated with the individual impurities and is
 independent of the separation of A and B. Using $g_{AA} = g_{BB}$, we find
\begin{equation} 
\label{equation:de1} \Delta E^S_{1} = \frac{2}{\pi} \text{Im}  \int dE f(E) \ln ( 1 - g_{AA} \lambda) \,,
\end{equation}
where the superscript $S$ refers to the substitutional impurities.
The second contribution is an interaction term dependent on their separation through the off-diagonal GF $g_{AB}$,
\begin{equation}
  \Delta E^S_{AB} = \frac{2}{\pi} \text{Im}  \int dE f(E) \ln ( 1 - \frac{g_{AB}^2 \lambda^2 }{(1-g_{AA}\lambda)^2}),
  \label{equation:deab}
\end{equation}
such that $\Delta E = 2 \Delta E^S_{1}  + \Delta E^S_{AB}$. It is easy to see that $\Delta E^S_1$ is both separation and configuration independent and takes a constant value.

\begin{figure}[t!]
\centering
\includegraphics[width = 0.45\textwidth]{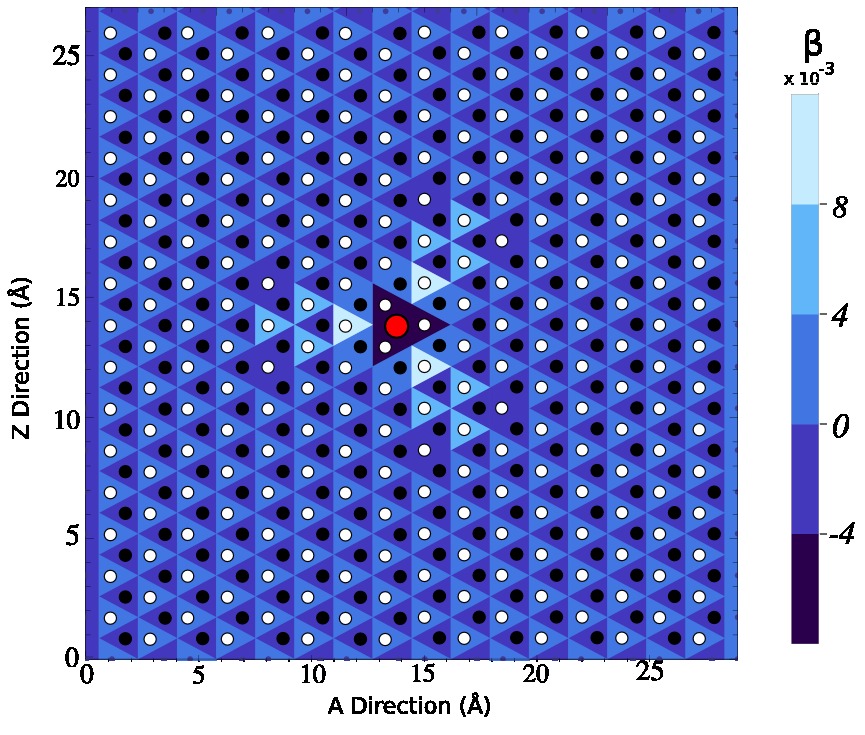}
\includegraphics[width = 0.45\textwidth]{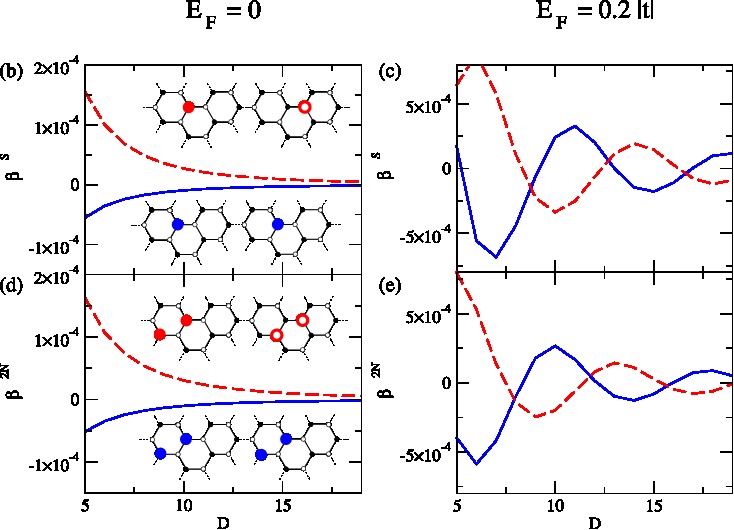}
\caption{(a) A substitutional nitrogen is moved around a central fixed nitrogen (red) and 
$\beta^S$ is plotted for each configuration at $E_F = 0$.
(b) Cross-section of plot (a) along the A-drection with the fixed impurity at $D=0$, the black (white) sublattice sites are indicated by the blue (red dashed) line
showing the $D^{-3}$ decay profile and the inset schematics show the configuration in both cases. 
(c) is a similar plot to (b) but with a Fermi Energy of $E_F = 0.2|t|$, where we see the presence of Friedel Oscillations and a decay of $D^{-2}$.
For two pairs of impurities we plot $\beta^{2N}$ as a function of $D$ in plot (d) for $E_F = 0$ and plot (e) for $E_F = 0.2|t|$ showing
the same features as for the case of two isolated impurities.}
\label{fig:nitrogens_armchair}
\end{figure}

To investigate the favourability of different configurations we define the dimensionless configuration energy function (CEF) for substitutional impurities \begin{equation}
\beta^S (E_F) = \frac{\Delta E_{AB}}{| 2 \Delta E_1  |}\,.
\label{beta_s}
 \end{equation}
This quantity describes the change in energy of the system due to the interference between the two impurities, relative to the total energy change in the system for two non-interacting (infinitely separated) impurities.
 Positive values of the CEF correspond to less favourable configurations whereas negative values correspond to favourable configurations which decrease the total energy of the system.
 By calculating the CEF for different impurity configurations we can establish which are energetically favourable and thus more likely to be realised in experiment. 
A map of $\beta^S (E_F=0)$ values for a large number of different configurations is shown in Fig. \ref{fig:nitrogens_armchair} (a).
 One impurity is fixed at the red circle corresponding to a site on the black sublattice. $\beta^S$ is then calculated with the second impurity located at each of the sites on the map, with the shading of the triangle surrounding each site corresponding to the $\beta^S$ value for that configuration. We note that, with the exception of nearest-neighbour site impurities, a general trend is seen where the second impurity prefers to locate on the same sublattice as the initial impurity. This trend gives rise to the chequerboard-like pattern seen in Fig. \ref{fig:nitrogens_armchair} (a) where same sublattice (black) sites are surrounded by darker triangles, corresponding to lower energy configurations, than the opposite sublattice (white) sites. It is also clearly visible in Fig. \ref{fig:nitrogens_armchair} (b) where we plot $\beta^S (E_F=0)$ for the two sublattices separately for armchair direction separations. We also note here that the magnitude of $\beta^S$ decays as $D^{-3}$.
 This is the same rate noted for FOs in $\Delta n$ for substitutional impurities and is easily explained by examining $g_{AB}$ and hence distance dependence of Eq. (\ref{equation:deab}) to first order in $\lambda$, which results in a similar equation to Eq. (\ref{equation:subs_fo}) for FOs in $\Delta n$. Indeed, the same plot for $E_F \ne 0$ in Fig. \ref{fig:nitrogens_armchair} (c) reveals an oscillatory behaviour and $D^{-2}$ decay rate, again matching the $\Delta n$ behaviour. Thus the FOs both in $\Delta n$ for a single substitutional impurity and in $\beta^S$ for a pair of impurities display the same distance dependent behaviour due to the similar dependence on off-diagonal Green functions that appears in both quantities.
 An important point to note is the discrepancy for the nearest neighbour impurity cases which Fig. \ref{fig:nitrogens_armchair} (a) shows to be most favourable. In this case the asymptotic behaviour extracted from our analytic expressions is evidently not yet valid. However we should also expect that our parameterization approach is not adequate to describe such impurities, as it treats the nitrogen impurities separately and neglects, for example, the additional overlap matrix elements required for two neighbouring nitrogen impurities. Another point to note is that the introduction of nitrogen impurities into graphene increases the total number of electrons in the system and leads to a shift in the Fermi energy. Thus for higher concentrations of nitrogen, the range for which same-sublattice doping is preferred is reduced due to the presence of the $\beta_S$ oscillations in Fig. \ref{fig:nitrogens_armchair} (c). However, the strength of the preference within this range may be increased by the slower rate of decay predicted for doped graphene.

\subsection{Impurity Segregation}
Recent experimental observations, corroborated by DFT calculations, seem to suggest that two nitrogen impurities in close proximity to each other prefer to occupy sites on the same sublattice in a quasi-neighbouring configuration\cite{nitrogen_original_science, nitrogen_second_experiment} as shown in Fig. \ref{fig:neighbours} (Left, Inset), hereon referred to as $N_2^{\bullet \bullet}$ configuration. This configuration is preferred over the nearest-neighbour configuration $N_2^{\bullet \circ}$ seen in Fig. \ref{fig:neighbours} (Right, Inset) which was calculated as being the most favourable configuration in the simple model above. Despite the limitations of the simple model for small separations, we note that the experimentally observed configuration fits the general trend of small separation, same sublattice configurations being the most favourable that the asymptotic behaviour of our model suggests. 

The real advantage to our FO-based approach to studying such systems becomes clear when we consider multiple $N_2^{\bullet \bullet}$ type impurities.
 Experimental evidence suggests that not only do pairs of nitrogen impurities prefer the $N_2^{\bullet \bullet}$ to the $N_2^{\bullet \circ}$ configuration,
 but that a pair of $N_2^{\bullet \bullet}$-like impurities prefer to locate on the same sublattice.
 In other words, that two $N_2^{\bullet \bullet}$ or two $N_2^{\circ \circ}$ impurities are formed in preference to one of each.
 This behaviour leads to domains in nitrogen-doped graphene with a large sublattice asymmetric doping.
 Such behaviour has been predicted to lead to interesting and useful transport properties\cite{nitrogen_doping_motivation}.
 Numerical investigation of such systems using DFT calculations is limited to small separations which makes it difficult to explore the behaviour which emerges in more highly-doped larger scale systems.
 By extending the model discussed for single substitutional N dopants in the previous section, we can use the Lloyd model to investigate these types of systems.
 In this setup, the individual impurities are now $N_2^{\bullet \bullet}$ or $N_2^{\circ \circ}$ defects, shown as insets in Fig \ref{fig:nitrogens_armchair} (e).
We consider a $N_2^{\bullet \bullet}$-type impurity at location A and introduce a second $N_2^{\bullet \bullet}$ or $N_2^{\circ \circ}$
 impurity a distance $D$ away at site B.
 We calculate the CEF for such a configuration analogously to Eq. \eqref{beta_s},
\begin{equation}
\beta^{2N} (E_F, D)= \frac{\Delta E^{N2}_{AB}}{| 2 \Delta E_1^{N2}  |} \,,
\end{equation}
where now $\Delta E_1^{N2}$ is the total change in energy in introducing a single $N_2^{\bullet \bullet}$ or $N_2^{\circ \circ}$ impurity.
 This quantity is plotted for the case when the second impurity is also a $N_2^{\bullet \bullet}$ (blue curve) and when it is a $N_2^{\circ \circ}$ (red dashed curve) in Fig. \ref{fig:nitrogens_armchair} (d). We note that, similar to the substitutional impurity case, same sublattice impurity configurations are preferential.

To benchmark our calculations, it is worth comparing our $\Delta E^{N2}_{AB}$ results to the DFT calculation performed in Lv et al. \cite{nitrogen_second_experiment_supplementary} for a single value of separation. In this work, calculations were performed for both two $N_2^{\bullet \bullet}$-type impurities and for a configuration with one $N_2^{\bullet \bullet}$ and one $N_2^{\circ \circ}$. In both cases the impurity pairs had a separation of approximately $D = 7$. An energy difference of 14meV is reported using the DFT calculation\cite{nitrogen_second_experiment_supplementary} compared to $2.3$ meV for the tight-binding model. In both cases the double $N_2^{\bullet \bullet}$  configuration was energetically favourable. The numerical discrepancy between the results is to be expected due to the overly simple paramterisation of the $N_2^{\bullet \bullet}$ impurity employed in the tight-binding model. However, the qualitative results for this model are not strongly affected by the local impurity parameterization, indicating also that the long-range sublattice ordered doping behaviour may not be unique to nitrogen. We emphasise that the same-sublattice configuration preference is noted for all separations in our model, explaining the long-ranged ordering seen in experiment\cite{nitrogen_original_science, nitrogen_second_experiment}. 

In a similar manner to that discussed for substitutional impurities, a finite concentration of $N_2^{\bullet \bullet}$ impurities shifts the Fermi energy away from the Dirac point and introduces oscillations in $\beta^{2N} (D)$. These oscillations, seen in Fig \ref{fig:nitrogens_armchair} (e) produce regions away from the initial $N_2^{\bullet \bullet}$ impurity  where a $N_2^{\circ \circ}$ is more favourable than a second $N_2^{\bullet \bullet}$ impurity. However, small increases in $E_F$ would preserve the same sublattice preference in local regions. This suggests that larger scaled N-doped systems may have alternating domains where each of the sublattices is dominant, and this prediction is consistent with experimental observations\cite{nitrogen_second_experiment}. An extension of the model discussed here to include a more accurate parameterization of the individual $N_2^{\bullet \bullet}$ impurities would provide a transparent and computationally efficient method to explore the formation and size of such domains, and to determine their dependence on the concentration of N dopants and the resultant Fermi energy shift.

\section{Conclusions} \label{sec:conclusion}

In this work we have derived analytic approximations for change in carrier density ($\Delta n$) and 
local density of states ($\Delta \rho$) in the high symmetry directions in graphene by employing a 
Stationary Phase Approximation of the lattice Green Functions.
We obtain excellent agreement with numerical calculations for single and double substitutional atoms,
vacancies, top adsorbed and bridge adsorbed impurities in the long-range limit, 
finding $\Delta n$ decays with distance ($D$) as $D^{-2}$ for all impurities for $E_F \neq 0$. 
At the Dirac Point, due to the disappearing density of states, 
we find the Friedel Oscillations in $\Delta n$ away from substitutional, top and bridge adsorbed impurities decay with as $D^{-3}$, but 
that in the case of vacancies and a resonant top adsorbed carbon $\Delta n$ is unchanged from the pristine case on all lattice sites
due to the symmetry of electron and hole states in the LDOS profile around the Dirac Point.
  In the case of top adosrbed carbon, which is less energetically favoured than the more naturally occuring bridge adsorbed carbon,
 the cross sublattice interference present in the bonding arrangement ensures $\Delta n$ does not vanish and can be seen through the Green Functions of the system.
\par \indent
Furthermore by expressing the total change in system energy due to the introduction of impurities through the lattice Green Functions
 we investigated how a sublattice asymmetry of both single and pairs of nitrogen dopants in graphene arises.
 We demonstrate that the dopant configuration energy is minimised where they share the same sublattice, 
a result which agrees with recent experiments \cite{nitrogen_original_science,nitrogen_doping_motivation,nitrogen_second_experiment} where such a distinct sublattice preference was found. 
\par \indent
It is possible for our method to be extended further by applying it to strained graphene,
 as has been done with the SPA approach to the RKKY\cite{group_strain_rkky} and the
behaviour of FOs in such a system has been studied theoretically only very recently\cite{dutreiz_friedel_strain}
 where a change in decay behaviour and sublattice asymmetry of 
the FOs due to the merging of the two inequivalent Dirac Points in the Brillouin Zone caused by inducing strain was found.

\section{Acknowledgements}

Authors J.L. and M.F. acknowledge financial support from the Programme for Research in Third Level Institutions (PRTLI).  
The Center for Nanostructured Graphene (CNG) is sponsored by the Danish National Research Foundation, Project No. DNRF58.

\bibliographystyle{ieeetr}
\bibliography{bibliography}{}

\end{document}